\documentclass[sigconf]{acmart}

\makeatletter
\renewcommand{\@authorfont}{\Large}
\renewcommand{\@affiliationfont}{\normalsize\normalfont}
\makeatother
\AtBeginDocument{%
  }

\copyrightyear{2026}
\acmYear{2026}
\setcopyright{cc}
\setcctype{by-nc-nd}
\acmConference[UIST '26]{The 39th Annual ACM Symposium on User Interface Software and Technology}{November 02--05, 2026}{Detroit, MI, USA}
\acmBooktitle{The 39th Annual ACM Symposium on User Interface Software and Technology (UIST '26), November 02--05, 2026, Detroit, MI, USA}
\acmDOI{10.1145/3830398.3830664}
\acmISBN{979-8-4007-2856-3/2026/11}

\usepackage{booktabs} 
\usepackage[ruled]{algorithm2e} 

\usepackage{graphicx}

\begin{document}

\title[
  \ourname: Geometry-Aware Microgesture towards Object-Agnostic Tangible Interaction
]{
  \ourname: Geometry-Aware Microgesture \revise{towards}
  Object-Agnostic Tangible Interaction
}



\author{Yinqiao Wang}
\orcid{0000-0002-6099-206X}
\authornotemark[1]
\affiliation{%
  \department{Department of Computer Science and Engineering}
  \institution{The Chinese University of Hong Kong}
  \city{Hong Kong}
  \country{China}}
\email{yqwang@cse.cuhk.edu.hk}

\author{Hao Xu}
\orcid{0000-0003-3676-5737}
\authornotemark[1]
\affiliation{%
  \department{Department of Computer Science and Engineering}
  \institution{The Chinese University of Hong Kong}
  \city{Hong Kong}
  \country{China}}
\email{xuhao@cse.cuhk.edu.hk}

\author{Qixuan Liu}
\orcid{0009-0001-5343-8052}
\authornotemark[1]
\affiliation{%
  \department{Department of Computer Science and Engineering}
  \institution{The Chinese University of Hong Kong}
  \city{Hong Kong}
  \country{China}}
\email{qxliu@cse.cuhk.edu.hk}

\author{Shengdong Zhao}
\orcid{0000-0001-7971-3107}
\authornotemark[2]
\affiliation{%
  \department{School of Creative Media}
  \institution{City University of Hong Kong}
  \city{Hong Kong}
  \country{China}}
\email{shengdong.zhao@cityu.edu.hk}

\author{Pheng Ann Heng}
\orcid{0000-0003-3055-5034}
\authornotemark[1]
\affiliation{%
  \department{Department of Computer Science and Engineering}
  \institution{The Chinese University of Hong Kong}
  \city{Hong Kong}
  \country{China}}
\email{pheng@cse.cuhk.edu.hk}

\author{Chi-Wing Fu}
\orcid{0000-0002-5238-593X}
\affiliation{%
  \department{Department of Computer Science and Engineering}
  \institution{The Chinese University of Hong Kong}
  \city{Hong Kong}
  \country{China}}
\email{cwfu@cse.cuhk.edu.hk}
\additionalaffiliation{%
\institution{Institute of Medical Intelligence and XR (IMIXR)}
\city{Hong Kong}
\country{China}}
\authornote{Corresponding authors.}

\renewcommand{\shortauthors}{Wang et al.}

\begin{abstract}
  This paper presents \ourname,  
\revise{an integrated}
framework 
\revise{towards} agnostic and tangible object interactions with microgestures.
Our goal is to support microgesture interactions across different everyday objects, \revise{with the capability to automatically leverage the geometric affordance of each object}.
We formulate a fingertip-aware detection pipeline to leverage generative 2D and 3D models for geometry enhancement and refinement.
We then introduce a usability-based method to prioritize the detected elements based on their ergonomic suitability for interactions. 
Building on this foundation, we further develop an AR system to transform everyday handheld objects into tangible user interfaces with 0D, 1D, and 2D microgesture interactions.
\revise{Across transitions among everyday cooking objects of varying shapes and sizes, \ourname~outperformed ablation baselines in task completion, usability (SUS), and workload (NASA-TLX).}
\revise{A further study with 10 objects demonstrates \ourname's generalizability across objects and grasps, highlighting its potential towards fluid, object-agnostic tangible interaction in real-world AR scenarios.}

\end{abstract}


\newcommand{\eg}{\emph{e.g.}\xspace} \newcommand{\Eg}{\emph{E.g.}\xspace}
\newcommand{\ie}{\emph{i.e.}\xspace} \newcommand{\Ie}{\emph{I.e.}\xspace}
\newcommand{\cf}{\emph{cf.}\xspace} \newcommand{\Cf}{\emph{Cf.}\xspace}
\newcommand{\etc}{\emph{etc.}\xspace} \newcommand{\vs}{\emph{vs.}\xspace}
\newcommand{\wrt}{w.r.t.\xspace} \newcommand{\dof}{d.o.f.\xspace}
\newcommand{\iid}{i.i.d.\xspace} \newcommand{\wolog}{w.l.o.g.\xspace}
\newcommand{\etal}{et al.\xspace}

\newcommand{\revise}[1]{#1}

\newcommand{\ourname}{ATOM\xspace}

\begin{teaserfigure}
\vspace{-4mm}
  \includegraphics[width=\textwidth]{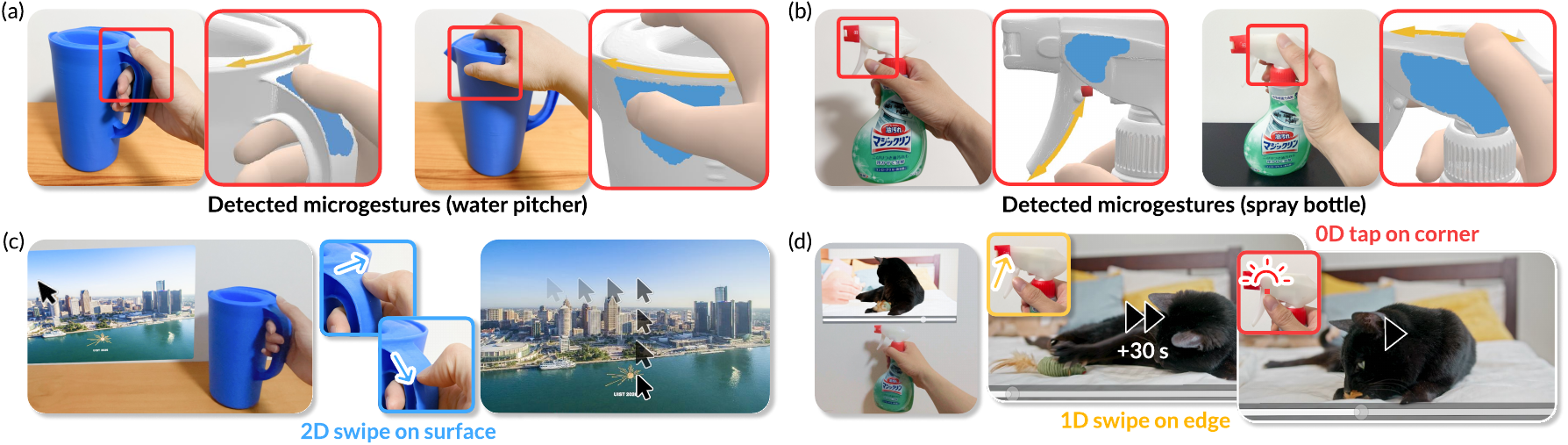}
  \vspace{-5mm}
  \caption{ 
Our method is able to detect corners (red), edges (yellow), and surfaces (blue) on different everyday objects and provide (a, b) tangible microgestures with different object grasping poses,~\eg, interacting with 
(c) the handle surface on a water pitcher to control cursor movement on the screen; 
and 
(d) the trigger of a spray bottle to control video playback.
More results on diverse grasping objects can be found in Figures~\ref{fig:qualitative} and~\ref{fig:application_scenarios}, and the supplemental video shows live captures of these interactive demos. 
}
  \label{fig:teaser}
\end{teaserfigure}


\maketitle

\section{Introduction}
Leveraging everyday objects as tangible props is 
an effective means to provide intuitive user interactions~\cite{tangiblebits97, insko2001passive, orit10, eva06grip, Aakar20replicate, saskia12embodied}.
Especially, everyday objects are not only ubiquitous and readily accessible but also offer natural affordance and tactile feedback, enabling pervasive and intuitive user controls.

Despite recent advances, existing solutions still have several limitations.
On the one hand, some earlier solutions are predicated on the geometric similarity between physical objects and their virtual counterparts, at object or part level, to ensure intuitive controls.
For example,~\cite{henderson08opportunistic, cheng10icon, funk14augment} preset 
specific objects as virtual controllers, such as manipulating a bottle's orientation to control the virtual rotation.
Subsequent solutions~\cite{hettiarachchi16annex, jain23ubitouch, zhou20grip} aim to automate the mapping between the virtual object to its physical object counterpart by considering their similarities in shape and affordance. 
However, the reliance on geometric similarity inherently limits the scalability of these solutions, 
as everyday objects and virtual controls are too diverse to consistently satisfy such assumptions.

On the other hand, some solutions~\cite{joshi19edge, niikura14anywhere, xiao13worldkit} relax this requirement by exploring 
geometric elements at smaller scales, such as edges and flat surfaces, on everyday objects.
He~\etal~\shortcite{he23ubiedge} 
detect and leverage sharp edges on objects for tangible interaction. However, their system requires the user to manually 
annotate the line segment on the physical object to author the tangible user interfaces. 
Furthermore, this approach is limited to edge-only interactions, which cannot provide inputs with higher degrees of freedom (DoFs).
%
Subsequently, He~\etal~\shortcite{he24adaptui} introduce a solution to automatically detect larger-scale geometric elements, including long straight edge segments and plane regions in the surrounding environment for quick TUI adaptation.
Yet, such environment-scale elements cannot be applied to everyday objects with curved edges and restricted surface areas (\eg, the trigger part of spray bottle in Figure~\ref{fig:teaser}).
To our knowledge, \revise{prior work has not yet presented a unified framework} for generalizable tangible interactions across diverse everyday objects.

To maximize the applicability
of everyday props, our key insight is to leverage their fine-grained local geometries,~\eg, protruding corners, sharp edges, and flat surfaces for richer types of interactions.
These ubiquitous elements can be found on a variety of everyday objects, without prior assumptions about object shapes and scales. 
Also, they provide distinct interaction metaphors driven by their unique tactile feedback, which naturally accommodates different interactions.
In this paper, we propose 
\revise{an integrated} framework, namely~\ourname (\textbf{A}gnostic and \textbf{T}angible \textbf{O}bject \textbf{M}icrogestures), to utilize everyday objects for tangible interactions by first detecting their fine-grained geometric elements 
%
then mapping these elements coherently to different finger-level actions based on the metaphors,~\ie, corner for 0D tapping, edge for 1D sliding, and surface patch for 2D swiping.
See examples in Figure~\ref{fig:teaser}.
With~\ourname, one can easily turn different everyday objects in hand into a tangible user interface that conforms to its local geometry and desired controls, by performing micro-interactions.

Locating such small-scale elements, however, is nontrivial. 
First, unlike straight edges and plane regions in~\cite{he24adaptui}, which have strong prior assumptions about object shape, corners, edge segments, and surface patches on everyday objects are difficult to find, due to their fine granularity.
This issue is further aggravated by the irregularity in object geometry and varying levels of detail, greatly affecting the detection accuracy.
%
Second, after detecting these geometric elements, not all of them are usable, given human interaction factors; \eg, one edge segment might not be preferable by the user due to an unnatural interaction pose.

To address these challenges, we first propose a fingertip-aware detection pipeline to locate geometric elements inside fingertip-reachable areas, thereby reducing the search space to a local region.
Further, to ensure generalizability to everyday objects, instead of relying on traditional algorithms~\cite{he23ubiedge} or deep models that are dedicated to indoor scenes~\cite{he24adaptui}, we exploit foundation models in computer vision for object understanding, which have shown strong cross-domain capabilities.
Specifically, leveraging foundational 3D and 2D generative models, we apply geometry enhancement both on 3D mesh and 2D edge maps to obtain regular and prominent features, thus providing robustness across objects with varying shapes and geometric noises. 
For the second challenge, we first conduct a comprehensive usability analysis of geometric elements for fingertip-based interaction, yielding a quantitative formulation. Guided by this formulation, a selection strategy is proposed to prioritize geometric elements with the highest ergonomics and accessibility.
Built on our framework, an interactive AR system is developed to detect geometry-aware microgestures and allow users to consistently perform finger-level tangible interaction across varying everyday objects, with intuitive control over multiple DoFs.
In summary, our contributions are as follows:
\begin{itemize}
    \item We introduce~\ourname, \revise{an integrated framework that maps fine-grained geometric primitives to microgesture vocabulary, towards object-agnostic tangible interactions.} 
    \item 
    Technically, we carefully design a fingertip-aware detection pipeline for geometric elements, equipped with 3D generative geometry enhancement and 2D generative geometry refinement for an accurate and robust element detection. 
    A comprehensive set of usability factors is proposed for prioritizing detected elements, ensuring their ergonomics and practicality in microgesture interaction.
    \item Based on our framework, we build an interactive AR system that enables tangible control using detected microgestures on diverse everyday objects. 
    We then compare our system with ablation baselines, and achieve the best performance in task completion, usability, and workload, showing the effectiveness of our framework design.
\end{itemize}

\section{Related Work}

\paragraph{Interaction using tangible props}
Various works have explored using physical objects as tangible props in VR/AR.
Very often, they leverage the geometric similarity between a physical object and a virtual object counterpart, typically by considering the whole object or a specific object part,~\eg, pressing the trigger of a spray bottle emulates the firing of a virtual gun.
Henderson and Feiner~\shortcite{henderson08opportunistic} 
first propose the idea of opportunistic control, aiming to map physical-object affordance to virtual widgets with similar interaction gestures. 
Yet, their prototypes are designed for specific objects.

Later, Cheng~\etal~\cite{cheng10icon}, Corsten~\etal~\cite{corsten13instant}, and Funk~\etal~\cite{funk14augment} generalize this idea 
to everyday objects.
Yet, users are required to explicitly specify suitable physical objects as tangible props for controlling virtual functions.
%
Given a virtual object, Hettiarachchi and Wigdor~\cite{hettiarachchi16annex} 
automate the process of identifying which nearby physical object can most closely match the virtual object, based on user preferences on object shape, size, and related properties.
Beyond considering object attributes, Jain~\etal~\cite{jain23ubitouch} additionally consider the similarity of hand gestures and contact points.
%
Instead of virtual-to-physical object mapping, Zhou~\etal~\shortcite{zhou20grip} inversely match physical objects to virtual primitive templates based on hand grip poses. 
More recently, Fan~\etal~\shortcite{tangiar25xu} propose a markerless tangible input system that tracks everyday objects in AR, enabling them to serve as passive controllers and virtual proxies with predefined behaviors.
\revise{On the other hand,
Monteiro~\etal~\shortcite{monteiro23} introduce a tool for authoring tangible AR applications in 
an interactive machine-teaching model.
Overall, prior approaches either rely on geometric similarity between physical and virtual objects at object/part level or require users to specify the interaction mappings per object, limiting their scalability across diverse everyday objects.
}

Another line of work explores low-level geometric elements (\eg, edges and surfaces) on everyday objects for tangible interaction, relaxing the requirement on object/part-level similarity.
For example,~\cite{xiao13worldkit, niikura14anywhere, xiao17support, han23blend, gil25prop} utilize flat (or nearly flat) surfaces on physical objects to host virtual interfaces and serve as touch input areas.
\revise{
Xu~\etal~\shortcite{surfacexr26xu} further advance XR inputs on everyday surfaces by fusing hand pose with smartwatch IMU signals to improve touch tracking and hand-surface gesture recognition.
}
Beyond surface-based inputs, He~\etal~\cite{he23ubiedge} utilize sharp edges on everyday objects to 
provide sharp tactile feedback. 
Nevertheless, it requires the user to preset the line segment as an affordance of tangible props through touch annotation, which is tedious.
Moreover, relying solely on edge fundamentally limits the supported 
interactions. For example, it prevents any tangible input on smooth and organic-shaped objects, 
whose surface can inherently provide continuous 2D inputs.

Later, He~\etal~\shortcite{he24adaptui} eliminate the manual definition of segments
by directly detecting environment-scale elements, including straight edge segments and large plane regions, in the physical environment,~\eg, the side rim and front surface of a table/desktop. 
While effective for rapidly adapting user-authored tangible interfaces to new physical environments, this approach relies on large environment-scale geometric elements 
%
whose availability \revise{is} rarely guaranteed, as everyday objects are mainly composed of curved edges and restricted surface areas.

Our framework 
\revise{is designed towards}
automatic and generalizable tangible interactions using everyday objects,
\revise{mapping fine-grained and usable geometric elements, including corners, edges, and surfaces, to interactive user controls.}
These ubiquitous elements can be found on a variety of everyday objects, without prior assumptions about their sizes, shapes, and structures.
Also, they provide distinct metaphors due to different tactile feedback~\cite {Plaisier2009}, which naturally accommodate interactions with rich degrees of freedom,~\ie, corner for 0D tapping, edge for 1D sliding, and surface for 2D swiping, enabling flexible and intuitive interactions.

\paragraph{Microgestures on object}
Some works study microgestures on objects, similar to the interactions introduced in our work. Their goals and methodologies differ significantly, which we briefly review below.
One line of work primarily investigates robust recognition of microgestures in the context of grasping objects, based on various sensing technologies~\cite{sharma21solofinger, rudolph22sense, sharma23sparse, kim23bio, lee25grab}.
Another line of research explores adaptive design for microgestures on grasping objects.
Sharma~\etal~\shortcite{sharma19} conduct an elicitation study to categorize microgestures performed on various representative objects. 
%
Joshi~\etal~\shortcite{joshi23transfer} create a design space at varying hand pose constraints, for transferring microgestures across different fingers and grasping objects.
To enable seamless transitions, Sharma~\etal~\cite{sharma24grasp} further explore gestures by dividing the grasping process into distinct phases.
More recently, Aponte~\etal~\shortcite{aponte24grav} design near-hand UIs for finger interactions by calculating the finger reachability.
Building on this, Caetano~\etal~\shortcite{caetano25graspr} propose an automatic approach to find the user-preferred UI locations during object grasps.
\revise{Beyond limiting gestures to reachable regions, \ourname automatically finds local geometric primitives with associated microgestures.}
Additional related work on 3D modeling can be found in~\cite{zhu2024pcf,zhu2025rethinking,zhueps3d,hu2024_cnsedit,hu2024neural,hu2026pegasus,liu2023exim,liu2026imagine,yan2026comp,li2026codrawagents,wu2026phymix,du2025hierarchical,hui2022neural,feng2026skelgen4d,feng2025wonderverse,hu2026cns,hui2022neuraltmp,zhu2026cos3d,zhu2021adafit,zhu2024ssp, Xu_2023_CVPR, Wang_2024_aaai, Xu_2024_CVPR, xu2025handboosterplus, Xu_2025, Wang_2025_CVPR, xie2026egohandicl, xu2026choir}.

\section{Design Requirements}
\begin{figure}[t]
    \includegraphics[width=\linewidth]{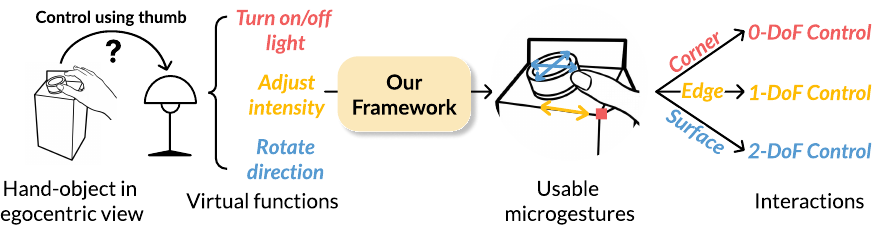}
    \caption{ Illustrating our workflow.}
    \vspace{-4mm}
    \label{fig:definition}
\end{figure}
Figure~\ref{fig:definition} illustrates the workflow.
Given a physical everyday object in the egocentric view, when the hand moves close to it, 
our framework first identifies the most usable local geometric elements,~\ie, its corner, edge, and surface, that are visible from the current view. These elements afford single-finger movements representing actions with varying degrees of freedom (DoFs), to control corresponding virtual functions.
For example, to control the virtual lamp using a milk carton, the user can use their thumb to tap its corner to turn on/off the light, or slide along its rim to adjust brightness, or swipe on its cover to control the light direction with 2 DoFs.

To achieve our goal, we consider the following requirements for designing our framework to provide users with optimal experiences:
\begin{itemize}
    \item \textbf{R1: Fine-grained detection}: The detected elements need to be fine-grained enough to offer different micro interactions within a local area. 
    \item \textbf{R2: User-suitable gesture}: Gestures on these geometric elements should be usable, considering their accessibility, ergonomics, interaction quality,~\etc
    \item \textbf{R3: Interactive and intuitive control}: To ensure a smooth and natural interaction, the system needs to provide real-time control that intuitively aligns with the user's intent. 
    
\end{itemize}

\section{\ourname{} Framework}
Our framework involves four steps. First, estimating and tracking hand and object pose (Section~\ref{sec:hand_object_pose_tracking});
Second, detecting available geometric elements within the finger's reachable region (Section~\ref{sec:feature_detection});
Next, selecting the elements with the highest usability (Section~\ref{sec:usablity});
\revise{Last, leveraging these elements for interactive control (Section~\ref{sec:interaction})}.

\begin{figure*}
    \includegraphics[width=\textwidth]{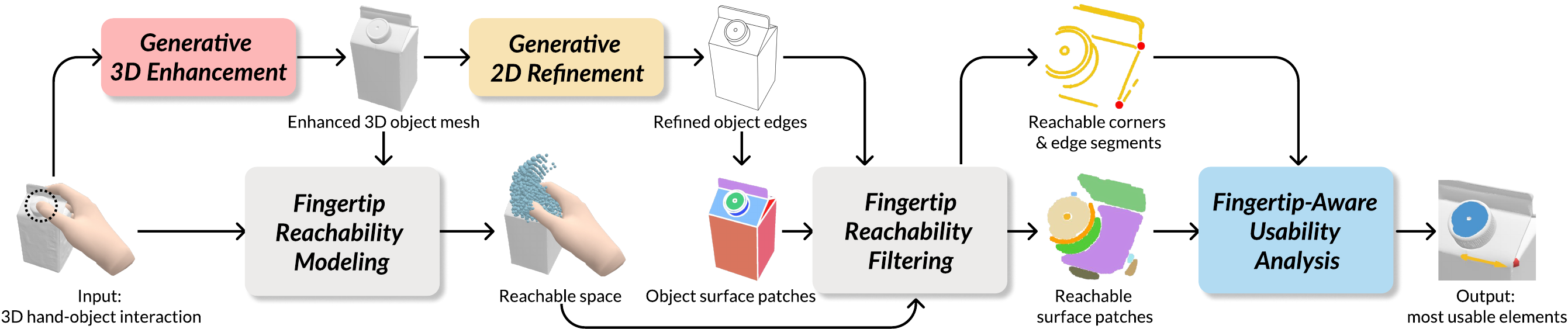}
    \caption{ The overview of our fingertip-aware detection pipeline.}
    \label{fig:pipeline}
\end{figure*}

\subsection{Hand-Object Pose Estimation and Tracking}
\label{sec:hand_object_pose_tracking}
We first estimate the 3D hand and object poses from egocentric-view images. 
Specifically, we adopt Unity's XRHand package~\cite{unity_xrhands_2023} to estimate the 3D hand joints. 
The hand is first converted to the MANO model~\cite{embodied17}, which is a parameterized hand representation that is convenient to use in subsequent computations.
Specifically, we achieve the conversion by minimizing the L2 distance between the raw joint estimates and the converted MANO joints. 
To estimate the 6D pose of the object, the object detection and tracking provided by ARKit~\cite{apple_arkit_2024} are used in our system.

\subsection{Fingertip-Aware Geometric Element Detection}
\label{sec:feature_detection}
\paragraph{Fingertip-aware detection pipeline}
The detection pipeline is shown in Figure~\ref{fig:pipeline}.
Given a 3D hand and object as input, we aim to identify the most usable geometric elements that can be reached by a certain finger.
To achieve this goal, we first extract the most basic geometry representation - edge map from the estimated 3D object.
%
%
Since the edge detection is a \revise{better-studied} problem in 2D vision, we adopt a 2D vision-based detection algorithm, following~\cite{he23ubiedge}.
Specifically, the 3D object mesh is rendered to a normal map, to which we apply the Canny edge detector~\cite{canny86}.

Next, unlike manually defining line segments by user touch in~\cite{he23ubiedge}, our framework is designed to automatically segment the detected edges and surfaces into smaller segments/patches to meet the fine-grained requirement (R1). 
%
Leveraging the geometric property that closed 2D contours enclose surface regions, we first use the full edge map to segment surfaces, by finding all the connected components (separated by edges) on the object rendered mask.
However, segmenting the edges themselves into discrete lines is inevitably challenging due to the complex network of intersecting curves and junctions that make boundaries ambiguous.
Directly using line segment detectors like~\cite{he24adaptui} is infeasible for everyday objects that usually contain curved edges (\eg, the cap of the milk carton in Figure~\ref{fig:definition}).
To reduce the complexity of the problem, we first locate edges that are reachable by the fingertip (refer to the next paragraph in this subsection), which are naturally split into different segments considering the finger's reachability.
Then, we apply Harris's algorithm~\cite{Harris1988ACC} to detect corners from this local edge map, and utilize DBSCAN~\cite{dbscan96} for consolidation. The consolidated corners further act as endpoints to segment edges that are interconnected.
Consequently, all reachable edge segments and corners are detected. More details can be found in the supplementary material.

\paragraph{Fingertip reachability modeling \& filtering}
\label{para:reachability}
To identify elements that can be reached by the fingertip, we calculate its movement space by iteratively updating the finger's joint rotation and applying the forward kinematics through MANO, constrained by the joint ranges in~\cite{nielsen04}. 
Meanwhile, we aim to avoid inter-penetration between hand and object by computing voxel-based intersection; and hand self-penetration using the algorithm from~\cite{libigl}. Fingertip positions with penetrated poses are filtered.
To learn more details, please refer to the supplementary material.
Subsequently, only geometric elements (\ie, corners, edge segments, and surface patches) that lie within this movement space are considered reachable.

\begin{figure}
    \includegraphics[width=\linewidth]{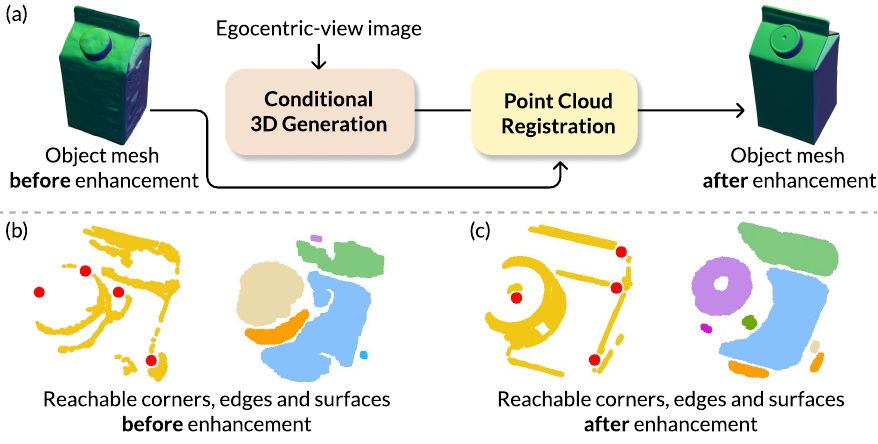}
    \caption{ Our generative 3D enhancement component. Corners are marked in red, segmented edges in yellow, and surface patches are distinctly color-coded.}
    \vspace{-4mm}
    \label{fig:generative_3d}
\end{figure}

Despite our carefully designed pipeline, we found that the detection performance can inevitably be inferior (Figure~\ref{fig:generative_3d}~(b)),
due to noises existing on both the 3D mesh and the 2D edge maps. 
First, the underlying 3D meshes often lack well-defined geometric structures, blurring critical corners and edges. Further, the 2D edge maps extracted from these imperfect 3D geometries naturally inherit this noise, suffering from discontinuities and background clutter.
Therefore, we propose to enhance the object geometry from the perspective of both 3D and 2D.

\paragraph{Generative 3D geometry enhancement}
To meet the requirement of fine-grained detection (R1), the object mesh needs to be of high quality, preserving prominent corners, sharp edges, and flat surfaces, \ie, with well-defined geometry, such that these elements can be correctly detected.
Despite recent progress in 3D reconstruction~\cite{kerbl3Dgaussians, mildenhall2020nerf}, the mesh processed by \revise{Marching Cubes}~\cite{marching87} always produces noise, thus hindering the edge detection.

Recently, conditional 3D generation models have shown strong capabilities of generating high-fidelity and plausible 3D geometry from a single-view image.
Hence, we propose generative 3D geometry enhancement by feeding the egocentric-view image from Section~\ref{sec:hand_object_pose_tracking} into the generative model~\cite{lai2025hunyuan3d25highfidelity3d} to obtain the 3D mesh with enhanced geometry, see Figure~\ref{fig:generative_3d} (a).
Since the generated object is not in metric scale and is unposed, we additionally perform point cloud registration between the output and the originally estimated object mesh. 
Considering generalizability, we first utilize a pretrained model~\cite{qin2022geometric} to roughly align two objects, and then use Iterative Closest Point~\cite{icp92} to achieve relatively precise alignment.
Figure~\ref{fig:generative_3d} illustrates the impact of our generative 3D enhancement module (b \vs c). By generating cleaner and sharper edge maps, \revise{it} greatly improves the subsequent geometric detection accuracy.

\paragraph{Generative 2D geometry refinement}
\begin{figure}
    \includegraphics[width=\linewidth]{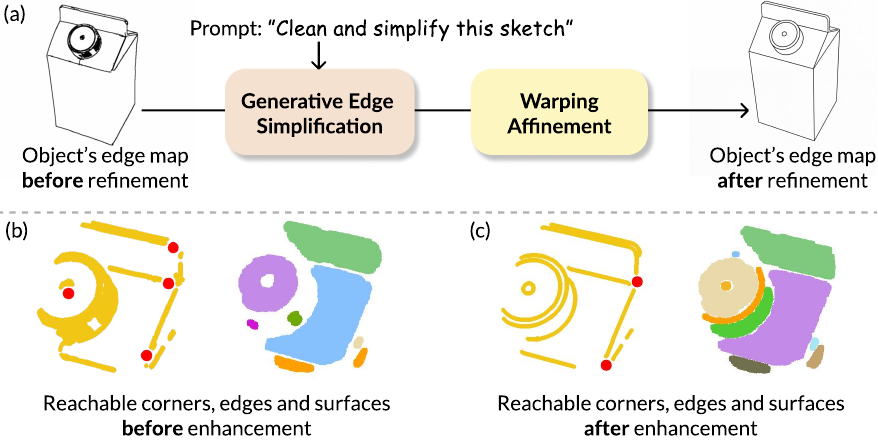}
    \caption{ Our generative 2D enhancement component.}
    \vspace{-4mm}
    \label{fig:generative_2d}
\end{figure}
Even with our generative 3D enhancement, the detection performance is still suboptimal, due to noise in the 2D edge map (see the left-most side of Figure~\ref{fig:generative_2d} (a)).
This is because vision-based edge detection algorithms are sensitive to local geometry variations, such as details around the cap of the milk carton,
leading to false detections. 

We propose to only capture prominent geometric characteristics that form the 2D object shape. 
To achieve this goal, we aim to extract a simplified representation of the rough edge map, which involves cleaning up less important lines and also repairing missing ones.
To this end, we leverage the foundation image generation model from~\cite{geminiteam2025geminifamilyhighlycapable} that has shown powerful capability for universal image editing tasks.
Specifically, we ask the model to perform sketch simplification on the input edge map, based on our prompt, to capture its prominent geometric features (Figure~\ref{fig:generative_2d} (a)).

Despite being an abstract form, the refined edge map is not guaranteed to be aligned with the original one, which is a common issue in current image editing models.
%
This can lead to the displacement of edges in the 3D space, reducing the detection accuracy.
To ensure the refined edge map closely aligns with the input, we perform feature matching between the original and refined edge maps using ORB descriptors~\cite{orb11} and estimate an affine transformation via RANSAC. Finally, we warp the generated image using this affine matrix to achieve spatially consistent results.
%
Our generative 2D geometry refinement further improves the detection accuracy of geometric elements, see Figure~\ref{fig:generative_2d} (b)~\vs(c).

\subsection{Fingertip-Aware Usability Analysis}
\label{sec:usablity}
\begin{figure}
    \includegraphics[width=\linewidth]{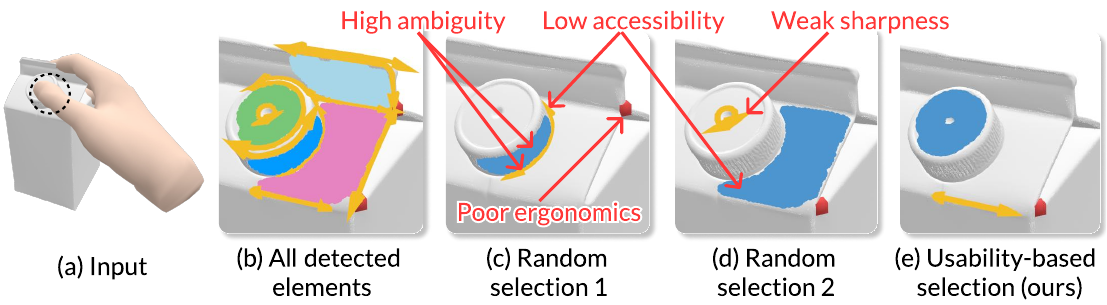}
    \caption{ Our usability analysis. }
    \vspace{-4mm}
    \label{fig:usability}
\end{figure}

In the previous step, we detected all the available geometric elements that can be reached by the fingertip. 
Yet, there might exist multiple different elements (Figure~\ref{fig:usability} (a)), due to the intricacy of the local object geometry. 
%
It remains unclear which ones are appropriate for microgesture interaction.
According to R2, in practice, users prefer gestures that are easier to use than those requiring more effort and with less interaction accuracy.
To this end, we consider a series of usability factors and formulate their mathematical representation for automatic gesture selection.

\paragraph{Usability Factors}
Given detection results, we aim to analyze the usability for each geometric element $\mathbf{E}_{i}$
to find the most suitable ones for user interaction, leveraging information from both the hand and the object.
Considering our design requirement R2, we propose the following usability scores: 
    
    \textbf{Reaching cost}: whether the cost of reaching the finger to the location is low, to reduce work. Formally, the score $\mathcal{S}_{\text{rch}}(\mathbf{E}_i)$ is defined as the L2 distance between the initial position of the fingertip and the center position of the element $\mathbf{E}_i$.
    
    \textbf{Fingertip accessibility}: whether the location is easy to access by the user's fingertip without being blocked by other object parts. 
    To measure the accessibility of a geometric element, we put hemispheres across its area and randomly emit rays to see the proportion intersecting with the object itself, which is formally defined as:
    \begin{equation}
        \small
        \mathcal{S}_{\text{acc}}(\mathbf{E}_i) = \sum_{i}^{N} \sum_{j}^{M} \operatorname{Raycasting}(p_{i}, \operatorname{SampleDirection}(j), O),
    \end{equation}
    where $N$ is the number of points on $\mathbf{E}_i$, $M$ is the number of rays, and $O$ is the triangle mesh of the object.
    Examples of poorly accessible edges and surfaces are shown in Figure~\ref{fig:usability} (c) and (d). 
    Particularly, although the surface area in (d) is slightly larger than that in (e), horizontal fingertip movement is consistently blocked by the cap.
    
    \textbf{Movement ergonomics}: whether the gesture is natural for a human's hand to perform. 
    We consider two aspects: first, the ergonomics of the whole hand pose; and second, the ergonomics of the fingertip.
    For whole hand pose, we adopt the twist-splay-bend frame for MANO~\cite{yang2021cpf}, which effectively 
    identifies abnormal poses. Specifically, our ergonomics score on hand $\mathcal{S}_\text{erg\_hand}$ is defined by projecting the rotation axis on each joint of the MANO hand to three independent axes, and computing the penalization of the abnormal axial components (refer to~\cite{yang2021cpf} for more details).
For the local fingertip $\mathbf{p}_{\text{finger}}$, we consider whether it can naturally approach the specified point on the object $\mathbf{p}_{\text{obj}}$ by evaluating the contact orthogonality - the alignment between the fingertip normal $\mathbf{n}_{\text{finger}}$, the point normal $\mathbf{n}_{\text{obj}}$, and their connecting vectors. 
For example, it is counterintuitive to use the fingernails instead of the fingerpad for precise contact with a specified point.
This alignment is formally defined as:
\begin{equation}
    \small
    \begin{aligned}
    \mathcal{S}_{\text{erg\_finger}} = & 
    \arccos \left( 
    \frac{ (\mathbf{p}_{\text{finger}} - \mathbf{p}_{\text{obj}}) \cdot \mathbf{n}_{\text{obj}} }
         { \|\mathbf{p}_{\text{finger}} - \mathbf{p}_{\text{obj}}\| \, \|\mathbf{n}_{\text{obj}}\| } 
    \right)
    + \\
    & \arccos \left( 
    \frac{ (\mathbf{p}_{\text{obj}} - \mathbf{p}_{\text{finger}}) \cdot \mathbf{n}_{\text{finger}} }
         { \|\mathbf{p}_{\text{obj}} - \mathbf{p}_{\text{finger}}\| \, \|\mathbf{n}_{\text{finger}}\| } 
    \right).
    \end{aligned}
\end{equation}
The final ergonomics score $\mathcal{S}_{\text{erg}}(\mathbf{E}_{i})$ for an element is obtained by calculating the sum of the hand and fingertip ergonomics scores ($\mathcal{S}_{\text{erg\_hand}}$ and $\mathcal{S}_{\text{erg\_finger}}$) at each point, and averaging this sum across all points within the element $\mathbf{E}_{i}$.
A counterexample of ergonomics is shown in Figure~\ref{fig:usability} (c), where the corner is difficult to touch with the thumb, given the grasping poses in Figure~\ref{fig:usability} (a).

    \textbf{Gesture ambiguity}: whether the gesture could trigger an unintended operation, due to the spatial proximity of the element and others. For example, in the case of a mug handle, the lateral curves and their adjacent surfaces are often close to each other.
    To this end, we calculate the Euclidean distances between the element and all other detected elements:
    \begin{equation}
    \small
\mathcal{S}_{\text{abg}}(\mathbf{E}_i) 
= \sum_{j\neq i} d_{\min}(\mathbf{E}_i, \mathbf{E}_j),
\end{equation}
  where $d_{\min}(\cdot, \cdot)$ denotes the closest distance between two point sets:
    \begin{equation}
    \small
    d_{\min}(\mathbf{E}_i, \mathbf{E}_j)
    =
    \min_{\mathbf{x} \in \mathbf{E}_i,\, \mathbf{y} \in \mathbf{E}_j}
    \left\| \mathbf{x} - \mathbf{y} \right\|_2 .
    \end{equation}
    Figure~\ref{fig:usability} (c) gives an example: the edge and surface around the cap are too close to each other, which easily leads to accidental mistouches during 1D and 2D interactions. 
    
    \textbf {Sharpness feedback}: whether the geometric corner/edge is acute enough to provide sharp tactile feelings.
    We aim to avoid mis-detected corners and edges, which may exist on occluding contours or geometric boundaries that are relatively smooth (~\eg, the one on bottle cap in Figure~\ref{fig:usability} (d)), which fail to serve as a physical landmark, as it is originally designed to~\cite{Plaisier2009, joshi19edge, sungjune16, he23ubiedge}, bringing ambiguous meaning.
    Specifically, we quantify the geometric sharpness of a single point $\textbf{p}$ by computing the average angles between its normal vector and those of its nearest neighbors:
    \begin{equation}
    \small
    S_{\text{shp}}(\mathbf{p}) 
    = \frac{1}{|\mathcal{N}(\mathbf{p})|}
    \sum_{i \in \mathcal{N}(\mathbf{p})}
    \arccos \left(
    \frac{\mathbf{n}_p \cdot \mathbf{n}_i}
    {\|\mathbf{n}_p\| \, \|\mathbf{n}_i\|}
    \right).
\end{equation}
  The sharpness of an element $\mathcal{S}_{\text{shp}}(\mathbf{E}_i)$ is then defined as the mean of its point-wise sharpness values.
%
For a more detailed explanation of these factors, please refer to our supplementary material.
\paragraph{Element prioritization}
To guarantee ergonomic interaction (R2), we first 
discard geometric elements that are nearly infeasible to use as below, by applying hard thresholds to specific usability scores:

    (i) Elements that have extremely low accessibility, making interaction physically infeasible;
    
    (ii) Corners or edges that lack sufficient geometric sharpness. This filters out false-positive detections and ensures reliable sharpness perception;
    
    (iii) Points on edges or surfaces that are in close proximity to other geometric elements. This reduces gesture ambiguity and unintended activations caused by spatially adjacent elements (\eg, ``double edges" formed by two closely parallel edges);
    
    (iv) Additionally, edges that are excessively short or surfaces that are too small. This mitigates the ``fat finger'' effect when users attempt to perform sliding gestures within a limited area.

For the remaining unfiltered candidates, we holistically consider all the usability factors to prioritize the optimal ones.
Since the original scores have different ranges and optimal directions, we standardize them to a uniform [0,1] scale where higher values indicate better usability. 
We then use their weighted sum to compute the final usability score for ranking
\begin{equation}
\small
\begin{aligned}
     \mathcal{S}(\mathbf{E}_i) & =  w_{\text{rch}} \mathcal{S}_{\text{rch}}(\mathbf{E}_i) + w_{\text{acc}}\mathcal{S}_{\text{acc}}(\mathbf{E}_i) +  w_{\text{erg}}\mathcal{S}_{\text{erg}}(\mathbf{E}_i) \\ 
     & + w_{\text{abg}}\mathcal{S}_{\text{abg}}(\mathbf{E}_i) + w_{\text{shp}}\mathcal{S}_{\text{shp}}(\mathbf{E}_i),
\end{aligned}
\end{equation}
where $w_{*}$ are hyperparameters.
The elements are ranked according to the usability scores.
Finally, we select at most top-$K$ element(s) for each DoF interaction,~\ie, top-$K$ corners, top-$K$ edges, and top-$K$ surfaces, if detected. We set $K$ to 1 by default.

\subsection{Interactive and Intuitive Control}
\label{sec:interaction}
\paragraph{Element parametrization}
So far, the detected elements are merely a set of 3D points without higher-order geometric structure. 
To enable these elements for digital control, a parameterization is needed.
A corner corresponds to a single point, which naturally represents a binary state that can be used as a discrete input,~\eg, a button.
To use the edge points set for continuous 1D input, we first use a minimum spanning tree that connects all of its points, and locate two ends based on node degrees. This ensures that all points are connected with the shortest line segments, which behave like a 1D slider.
For the continuous 2D input using the surface points set,
we calculate the two basis vectors from the tangent plane of the surface's manifold center, with which all the points on the surface patch are transformed into a local coordinate system. It behaves like a touch pad.
After the above process, all the geometric elements are converted into local user interfaces with real-time response (R3), which are ready to be mapped to digital functions. 

\paragraph{Swiping direction alignment}
For interaction, a further challenge lies in the ambiguous directionality of the detected elements. 
Because the pipeline only extracts points, it does not define the interaction polarity — such as which end serves as the origin of a 1D slider.
%
This semantic gap can severely degrade the experience. For example, when fast-forwarding a video, users naturally expect a left-to-right or bottom-to-top swipe. An arbitrary or reversed mapping would violate these expectations and result in a highly counter-intuitive interaction.
%
To maximize the intuitiveness regarding swiping interaction (R3), we propose an automatic adjustment strategy to align the swiping direction with downstream applications.
Specifically, we first recognize the most aligned axis of the edge, based on the angle between it and the $x$, $y$, and $z$ axes in the user's camera coordinates. 
According to the aligned axis, the system automatically adjusts the edge to the positive direction of the aligned axis.
For example, 
\revise{during video playback},
the direction of the edge will be flipped if its aligned axis is $x$ but is along the negative direction (from right to left), or if the aligned axis is $y$ but is along from top to bottom.
Similar adjustment is applied to the two axes of the detected surface,
%
~\eg, aligning them to left-to-right and bottom-to-top directions in the \textit{Cursor Movement} scenario, to mimic the physical movement of a mouse to control the cursor on the \revise{screen}.
%
Note that this adjustment is application-dependent, as the input-output behavior of each application scenario might vary based on the user's preference.

\begin{figure*}
    \includegraphics[width=\textwidth]{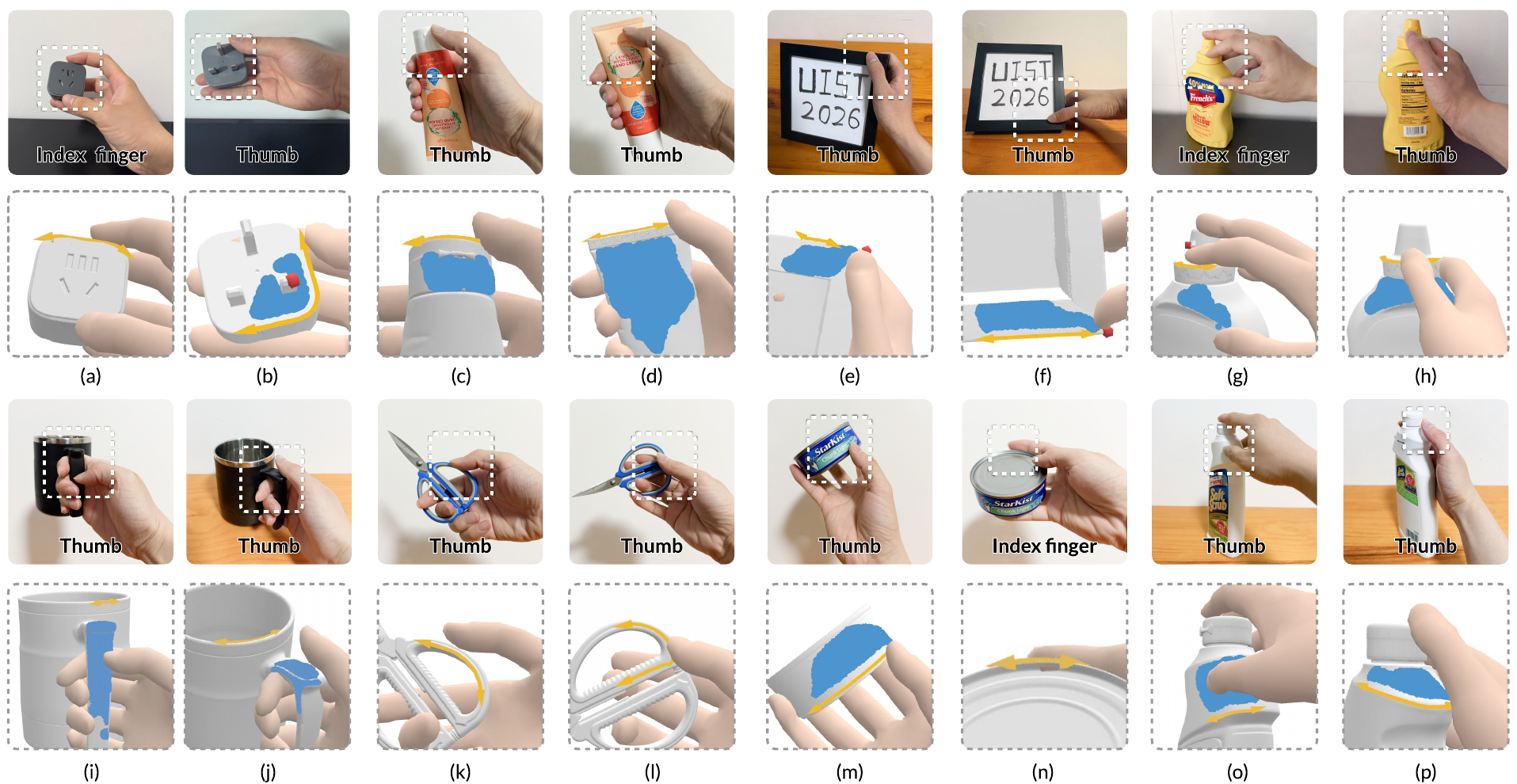}
    \caption{ Detected geometric elements (corners: red, edges: yellow, surfaces: blue) across various objects and grasping poses. The system adapts to the finger used for interaction (\eg, the index finger in \revise{(a, g, n)}), yielding distinct available interaction regions. 
    }
    \vspace{-3mm}
    \label{fig:qualitative}
\end{figure*}
\paragraph{Contact and active finger detection}
\revise{
Contact is detected as near-surface motion relative to parameterized geometric elements, using fingertip proximity and motion consistency; gesture start/end correspond to entering/leaving valid contact. The active finger is selected via a heuristic over proximity (<0.5 cm) and motion range (>2.5 cm/s). For simplicity, multi-finger movement is not currently supported. 
}
%
To enable fluent interaction (R3), we sample four contact points around the fingertip (the tip, the pad, and the left and right lateral sides), such that different areas of it can be utilized. We then adopt a voting mechanism with a temporal window size of 10 to determine which geometric element 
is contacted with.

\section{Results and Applications}
\paragraph{Implementation details}
Our system employs a client-server architecture. The client-side AR environment is implemented in Unity and experienced via an Apple Vision Pro headset. This headset communicates remotely with a Python-based server hosted on an Apple laptop (M2 Pro 3.49 GHz CPU, 32 GB RAM), which offers the capability of geometric element detection. 
\revise{More details about the setup can be found in the supplementary material}.

\revise{
\paragraph{Detected microgestures}
Figure~\ref{fig:qualitative} shows microgestures detected on eight different objects under various grasping poses.}
For each case, the top is the ego-centric view image,  with the active finger highlighted, and the bottom shows the most usable geometric elements detected by our framework (red cube for 0D control, yellow arrow for 1D control, and blue area for 2D control).
 (a): The adapter is grasped from the front side, and the user can use the index finger to slide along its rim for 1D control.
 (b): The same adapter is grasped from the back side. Now using the thumb, the user can additionally tap the pin for 0D control and also swipe on the small area for 2D control;
 (c) and (d): Grasping the hand cream from a different side. In both cases, the geometric edges and surfaces are successfully detected;
 (e) and (f): Putting the hand at different positions of the photo frame. The corners, edges, and surfaces are correctly detected for interaction with different DoFs;
 (g) and (h): When the user puts the index finger or the thumb near the cap of the mustard bottle, the most usable curved edges are detected, along with a spatially distant surface area that is finger ergonomic.
 Interestingly, our framework detects the corner from a very small protruding part of the cap in (g), which provides the user with an intuitive connection with a button.

\begin{figure*}
    \includegraphics[width=\linewidth]{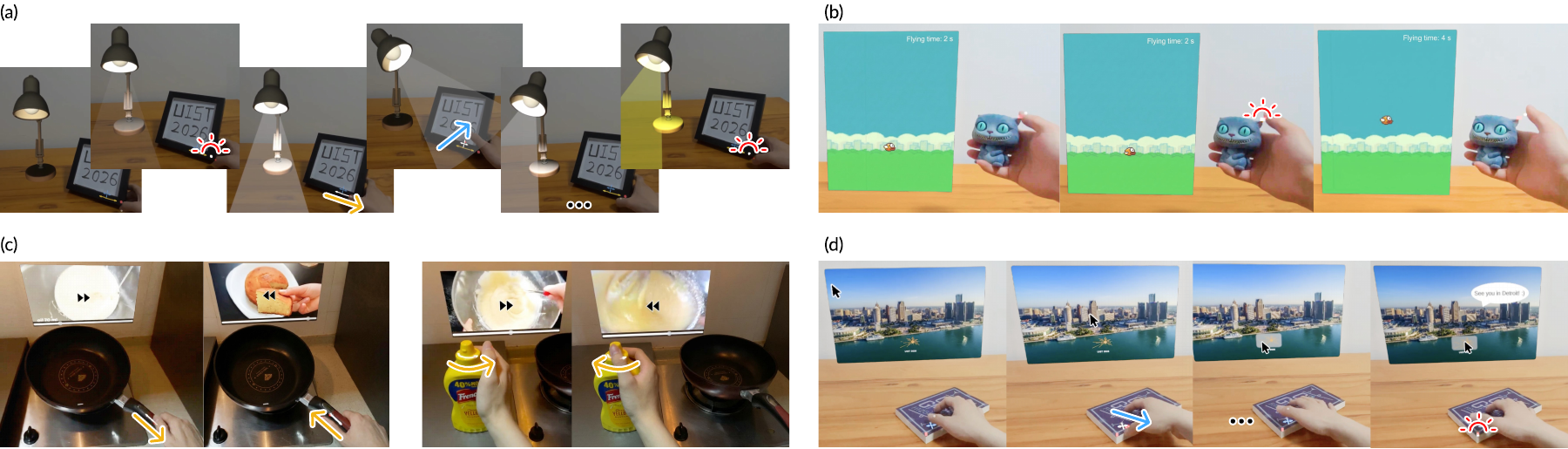}
    \vspace{-2mm}
    \caption{ Four live-captured AR applications, in scenarios of a living room (a, b, d) and kitchen (c).}
    \label{fig:application_scenarios}
\end{figure*}

\paragraph{Application scenarios}
In Figure~\ref{fig:application_scenarios}, we showcase four live-captured applications in our AR system to demonstrate its effectiveness.
(i) \textit{Lamp control}: We show the mixed use of controlling a virtual lamp using the photo frame: tapping corner (0D) to turn on/off the lamp, or switch the light color from white to yellow; sliding on edge (1D) to adjust the light intensity; and sliding on surface (2D) to rotate the light about the $x$- and $y$-axes, enabling continuous control of its orientation (Figure~\ref{fig:application_scenarios}~(a)).
(ii)\textit{Flappy Bird}: We use the corners detected at the ear of the toy cat to control a prototype Flappy Bird game, see Figure~\ref{fig:application_scenarios}~(b). To play the game, the user can touch the ear to send a flying-upward signal to the bird.
(iii)\textit{Video playback}: In the kitchen scenario (Figure~\ref{fig:application_scenarios}~(c)), the user uses the detected edge at the handle of a frying pan to control the playback of a video on cooking.
Once switched to grasping the mustard bottle, the thumb-based interaction naturally turns to sliding on the curved edges around its cap.
In both cases, the video plays forward each time the user swipes right, and backward otherwise.
(iv)\textit{Cursor movement}: In Figure~\ref{fig:application_scenarios}~(d), we use the detected surface on the book cover to control the movement of a cursor on a virtual \revise{screen}, and tap its corner to click the ``UIST 2026" icon.
The local x and y axes of the surface naturally map to the horizontal and vertical dimensions of the \revise{screen}. This enables the continuous recognition of arbitrary sliding directions, functioning like a touchpad.

\section{Evaluation}
We comprehensively evaluated \ourname through two user studies and a system-level analysis. Study~1 compared the full pipeline with three ablation baselines, while Study~2 tested generalizability across everyday objects and grasp poses. The system-level analysis assessed the performance of individual subsystems.

\subsection{User Study 1: Ablation}
\label{sec:ablation_study}
\subsubsection{Procedure}
\paragraph{Goals}
To validate our framework of geometry-aware microgesture detection and interaction, we conducted an experiment comparing our full system against ablation baselines, with particular emphasis on the following questions:
\textbf{Q1:} Does our full detection pipeline outperform ablation baselines in terms of detection accuracy and interaction performance?
\textbf{Q2:} Does the improved detection of geometric elements bring better task speed, accuracy, and overall system usability?
\textbf{Q3:} Can our system support consistent, fluid microgesture interaction when users switch between different physical objects during a realistic task?

\paragraph{Participants}
12 participants (3 female and 9 males), ranging from 20 to 30 years old, volunteered for the experiment. Among these, four had prior experience with VR/AR head-mounted displays.

\begin{figure}
    \includegraphics[width=\linewidth]{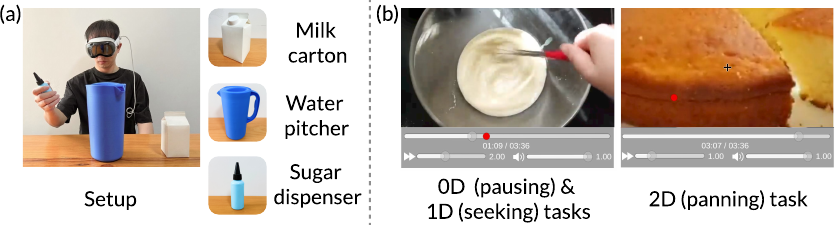}
    \vspace{-3mm}
    \caption{ User study setup.}
    \vspace{-2mm}
    \label{fig:userstudy_setup}
\end{figure}

\paragraph{Hardware and setups}
The experiment was conducted using an Apple Vision Pro mixed-reality headset. \revise{The system 
detects microgestures performed on physical objects and render the digital interface (a video player) directly in the user's field of view.} To represent common objects used in daily tasks, three physical props of varying shapes and sizes were used: a milk carton, a water pitcher, and a sugar dispenser (Figure~\ref{fig:userstudy_setup} (a)).

\paragraph{Tasks}
We designed a simulated kitchen scenario in which a user watches a cake-baking tutorial video (Figure~\ref{fig:userstudy_setup} (b)) while handling various physical objects. Because the user's hands are occupied holding these props, they must interact directly via the objects' geometry to control video playback.
To accurately reflect real-world viewing behaviors, we considered three essential operations used when learning from a tutorial video:
    (i) \textbf{Pausing:} Useful when the user identifies a point of interest or requires additional time to carefully inspect a specific step.
    (ii) \textbf{Seeking time:} Necessary when the user wishes to review previous footage to verify their physical actions, or skip ahead if they are already familiar with the current step.
    (iii) \textbf{Panning the focal region:} Crucial for observing intricate details. When the video is zoomed in, the user must be able to navigate the viewport to their specific region of interest.

We logically mapped these three operations to our system's three-degree-of-freedom (DoF) geometric controls: a 0D corner microgesture for pausing, a 1D edge microgesture for timeline seeking, and a 2D surface microgesture for panning the zoomed-in video. A trial occurred as follows:
    (i) \textbf{0D Task (Pausing):} A target timestamp was displayed as a red dot on the video progress bar. The participant was required to use a 0D corner microgesture to pause the playhead (white dot) as close to the red dot as possible. Once satisfied, the user clicked a confirm button with their other hand to end the trial. If the system failed to detect a corner and the participant could not trigger the pause action at all, they were allowed to click a ``skip'' button.
    (ii) \textbf{1D Task (Seeking):} A target timestamp was displayed as a red dot. The participant used a 1D edge microgesture to slide the playhead toward the target. Once the playhead was sufficiently close, the red dot turned green. The participant then had to release the sliding operation and hold the playhead stationary at the target for a brief dwell period, which the system registered as a success and automatically advanced to the next trial. If the action could not be completed, the user could skip the trial.
    (iii) \textbf{2D Task (Panning):} In a zoomed-in video state, a target location was displayed as a red point. The participant used a 2D surface microgesture to drag the video's focus point (a crosshair) to the red point. Similar to the 1D task, the target turned green when the crosshair was close enough, and the user had to release and dwell to complete the trial. Unsuccessful trials could be skipped. 

\paragraph{Experimental design}
We ablate core components from our \revise{full pipeline}, resulting in three baselines:
the generative 3D enhancement (ours w/o Gen 3D), generative 2D refinement (ours w/o Gen 2D), and the fingertip-aware usability analysis (ours w/o UA).

A within-participants design was used. Each participant tested four methods (our full pipeline and three ablation baselines) across the three physical objects (carton, pitcher, dispenser). The presentation order of the methods and objects was completely counterbalanced across the 12 participants using a Latin Square design.
For each method-object combination, the participant grasped the object following our suggested predefined poses to ensure that variations in grasping did not confound the results. They then performed the 0D, 1D, and 2D tasks sequentially. To avoid learning bias, participants completed four trials per task with randomized target values (e.g., different target timestamps or pan locations; exact values are detailed in the supplementary material).
A set of warm-up trials was provided before the formal experiment began to familiarize participants with the tasks, \revise{without telling them how to use the detected corners, edges, and surfaces}. In summary, the design was as follows (excluding warm-ups):
12 participants $\times$
4 methods (full pipeline + 3 baselines) $\times$
3 objects (carton, pitcher, dispenser) $\times$
3 tasks (0D, 1D, 2D) $\times$
4 trials per task = 1,728 total possible trials. 
\revise{Completed trial counts vary, 
as there are different numbers of corners/edges/surfaces for different objects and baselines.}

\subsubsection{Results}
\label{subsubsec:result}
\paragraph{Statistics of failed trials}
\begin{table}[t]
\caption{Statistics of failed trials.}
\resizebox{\linewidth}{!}{
    \begin{tabular}{lccc}
        \toprule
        Method & 0D Pausing Task $\downarrow$ & 1D Seeking Task $\downarrow$ & 2D Panning Task $\downarrow$ \\
        \midrule
        Ours w/o Gen 3D & 0.00\% & 4.17\% & 1.04\% \\
        Ours w/o Gen 2D & 14.58\% & 20.83\% & 11.46\% \\
        Ours w/o UA   & 4.17\% & 0.00\% & 69.79\% \\
        Ours (Full)     & 0.00\% & 0.00\% & 0.00\% \\
        \bottomrule
    \end{tabular}
    }
    \vspace{-4mm}
    \label{tab:skip_test_statistics}
\end{table}

We first count the number of failed trials due to the inaccessibility and difficult usability of detected microgestures in Table~\ref{tab:skip_test_statistics}. Our full pipeline consistently provides users with successful trials, while other methods inevitably find geometric elements that are either inaccessible or too hard to use.
\paragraph{Task performance}
Next, we analyze all continuous objective and subjective metrics (\ie, accuracy, completion time, and scores from Likert scales) using Repeated Measures ANOVA (RM-ANOVA). For all significant main effects, post-hoc pairwise comparisons were conducted with Bonferroni correction. 

\begin{figure}
    \includegraphics[width=\linewidth]{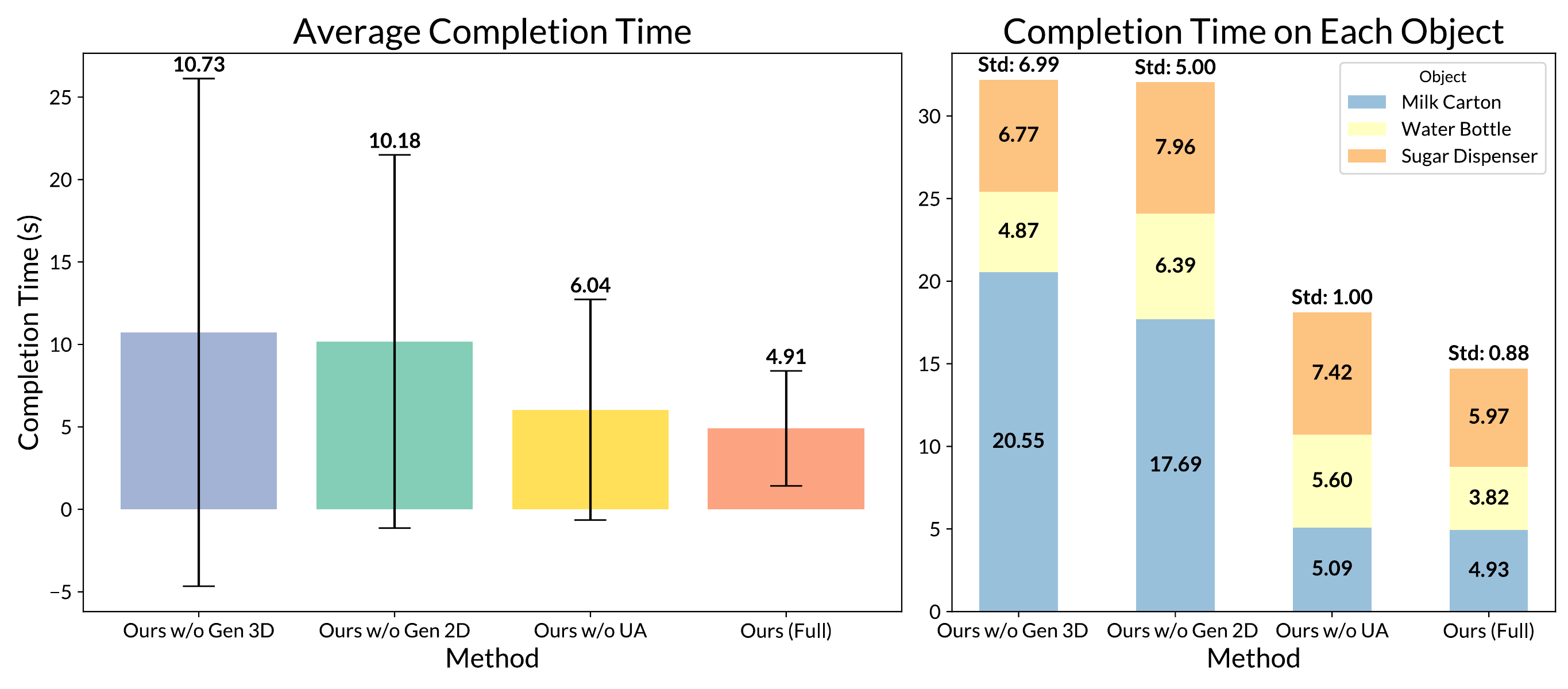}
    \caption{ Completion time of 1D task. }
    \vspace{-3mm}
    \label{fig:completion_time_1d}
\end{figure}
\begin{figure}
    \includegraphics[width=\linewidth]{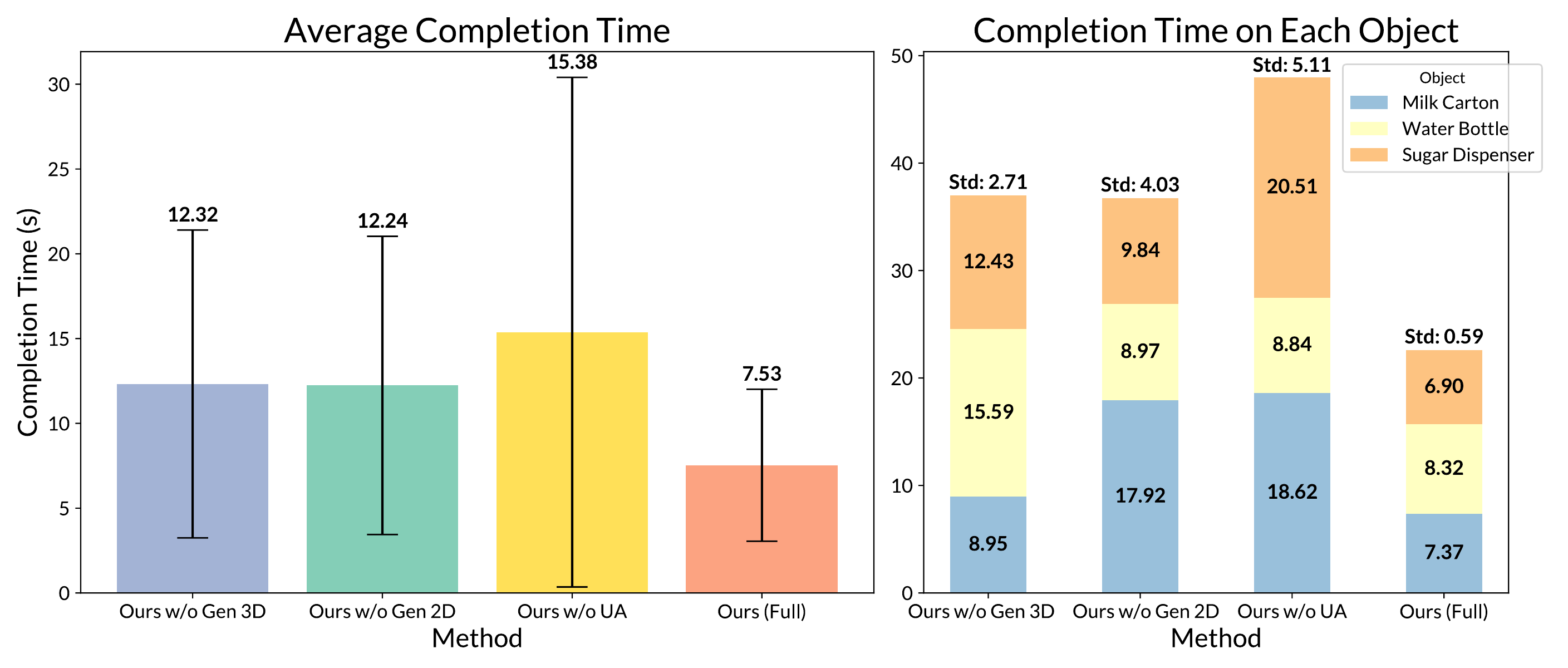}
    \caption{ Completion time of 2D task. }
    \vspace{-4mm}
    \label{fig:completion_time_2d}
\end{figure}
For the 0D control, accuracy was measured as 
\revise{one minus the absolute temporal deviation from the target,}
normalized by the total video length. If a user was forced to skip a trial due to detection failure, the accuracy for that trial was recorded as 0. 
Post-hoc tests show that our full pipeline achieved high accuracy ($M{=}98.41\%, SD{=}2.38$), comparable with the highest-performing baseline ours w/o Gen 3D ($M{=}98.15\%, SD{=}2.19$)($p{>}0.05$).
This comparable performance is primarily due to the spatial tolerance inherent to 0D pausing tasks:
during interaction, whether the corner is detected at the correct position (\ie, physical corner) makes no big difference for pausing at a correct time, as long as it remains reachable.
However, the Likert-scale results (Section~\ref{sec:subjective}) show that such ``false-positive" corners offer less meaningful tactile feedback and are less intuitive.

For 1D and 2D controls, completion time was measured from the moment the user first triggered the interaction to the moment the dwell-state success was registered. Skipped trials were treated as failures (timeouts) and excluded from the continuous time analysis. 
Figure~\ref{fig:completion_time_1d}~(a) shows that for the 1D task, our system enables users to slide to the target position in less than 5 seconds, which is comparable with other state-of-the-art tangible sliders~\cite{he23ubiedge, embodied20maxime, wall12yvonne}. 
Pairwise comparisons indicate that the full pipeline ($M{=}4.91$, $SD{=}3.49$) tended to outperform all baselines: ours w/o Gen 3D ($M{=}10.73$, $SD{=}15.39$; $t(11){=}-4.65$, $p{=}0.0036$), ours w/o Gen 2D ($M{=}10.18$, $SD{=}11.32$; $t(11){=}-8.30$, $p{=}0.0001$), and ours w/o UA ($M{=}6.04$, $SD{=}6.69$; $t(11){=}-1.37$, $p{=}0.2440$).
For the 2D task, our full pipeline ($M{=}7.53$, $SD{=}4.48$) significantly outperformed the best baseline, ours w/o Gen 2D ($M{=}12.24$, $SD{=}8.79$; $t(11){=}-7.80$, $p{=}0.0002$).

Additionally, we report the task completion time on different objects, see Figure~\ref{fig:completion_time_1d} (b) and Figure~\ref{fig:completion_time_2d} (b).
Our full pipeline achieves both the shortest completion time (4.93/3.82/5.97 for the 1D task and 7.37/8.32/6.90 for the 2D task) \revise{with} the lowest standard deviation between three objects (0.88 for the 1D task and 0.59 for the 2D task).

\paragraph{Subjective preference}
\label{sec:subjective}
To provide a comprehensive evaluation of user experience, participants completed a post-condition Likert scale questionnaire after testing each method. This included the System Usability Scale (SUS), the NASA Task Load Index (NASA-TLX), \revise{and five custom questions (5-point):}
    (i) \textbf{Detection Accuracy:} The detected elements are correctly found (\eg, corners are protruding, edges are sharp, and surfaces are flat on the physical object).
    (ii) \textbf{User Accessibility:} My finger can touch the detected elements without much effort (\eg, without obviously moving other fingers).
    (iii) \textbf{User Ergonomics:} I feel natural and comfortable when using these detected elements for interaction.
    (iv) \textbf{Intuitive Control:} The interaction result intuitively aligns with my intention.
    (v) \textbf{System Affordance:} Just by looking at the detected elements, I immediately understood how to use them for interaction.

\begin{figure}
    \includegraphics[width=\linewidth]{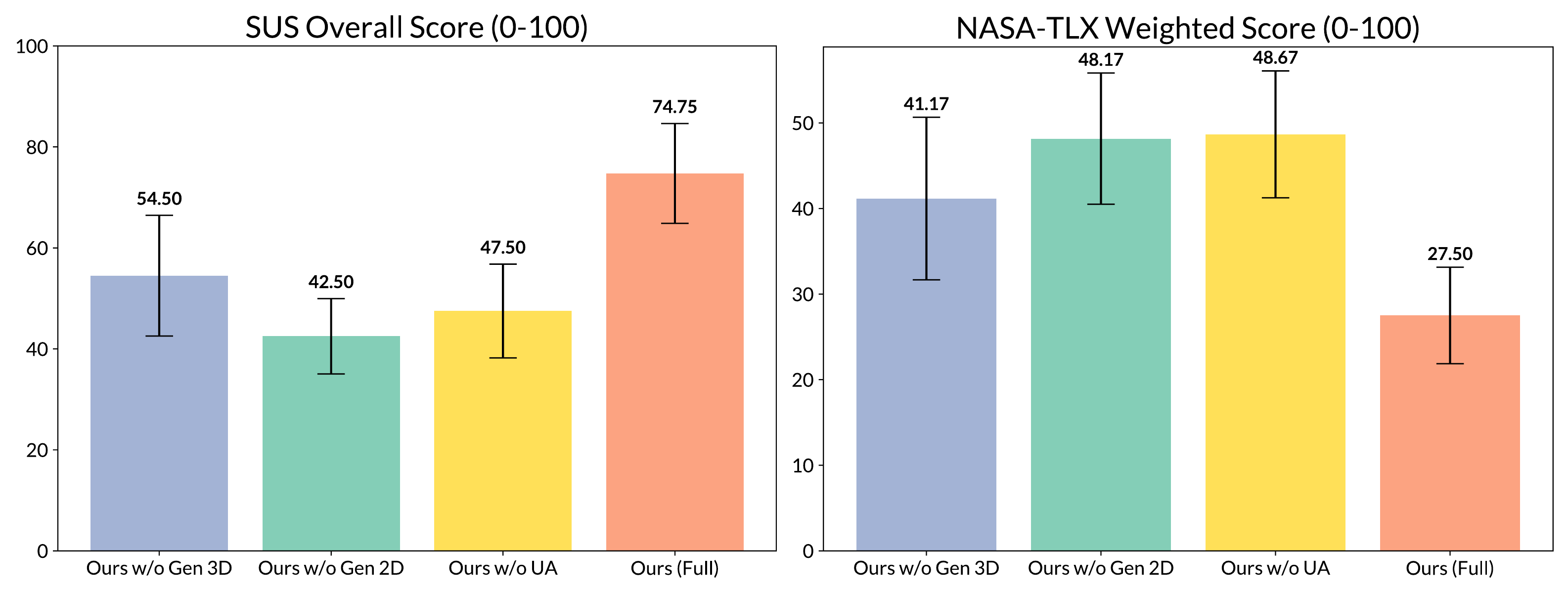}
    \caption{ SUS ($\uparrow$:better) and NASA TLX ($\downarrow$:better) scores.}
    \vspace{-3mm}
    \label{fig:user_study_qualitative}
\end{figure}
\begin{figure}
    \includegraphics[width=\linewidth]{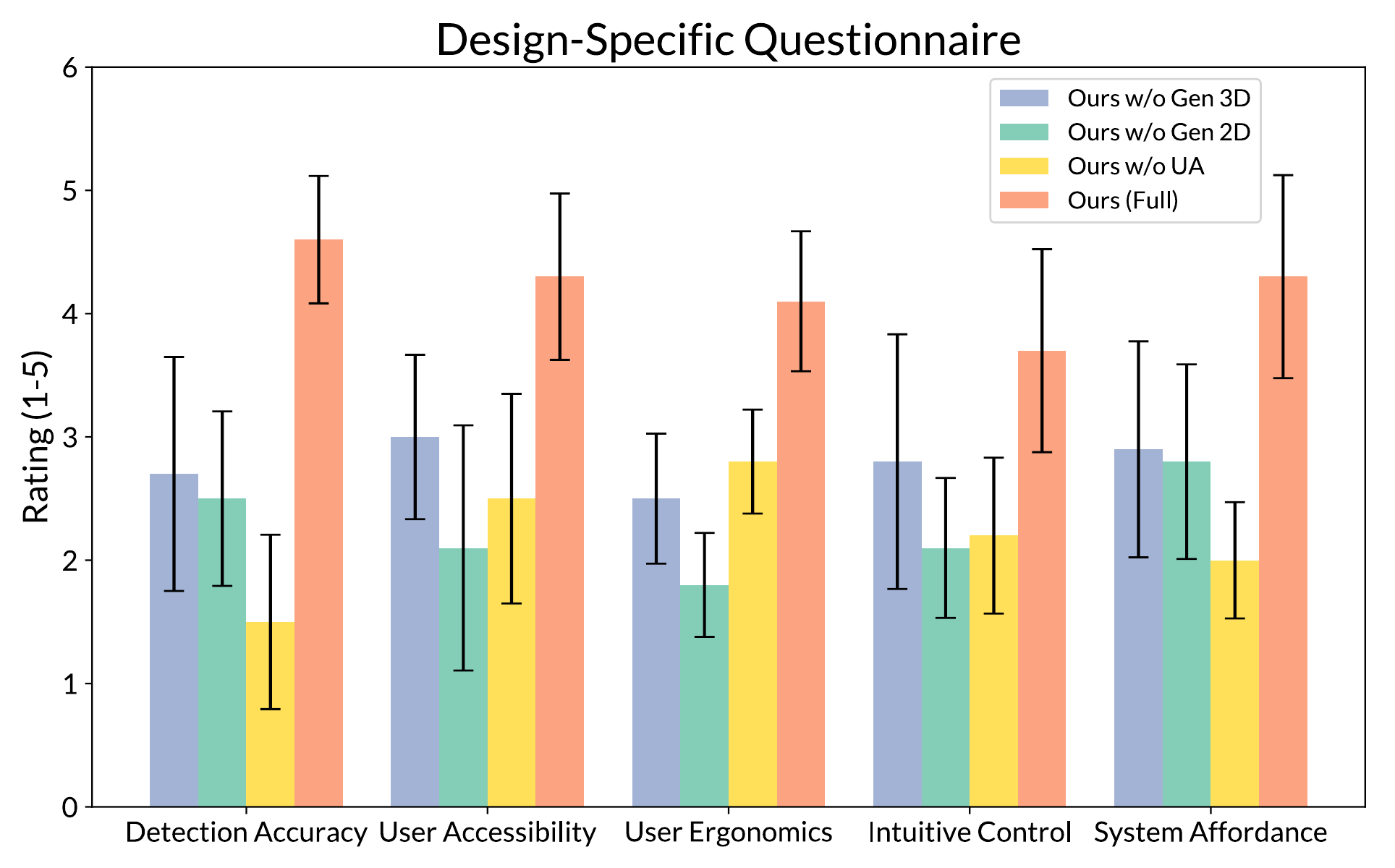}
    \caption{ Ratings of our additional Likert scale. }
    \vspace{-4mm}
    \label{fig:qualitative_additional}
\end{figure}

As shown in Figure~\ref{fig:user_study_qualitative}~(a), our full system achieved a much higher SUS score ($M{=}74.75, SD{=}9.89$) than the best baseline, ours w/o Gen 3D ($M{=}54.50, SD{=}11.95$)($t(11){=}3.30, p{=}0.0183$), indicating its superior overall usability. 
Furthermore, it yielded the lowest perceived workload (Figure~\ref{fig:user_study_qualitative} (b)) on the NASA-TLX ($M{=}27.50, SD{=}5.64$), which is significantly better than the best baseline ours w/o Gen 3D ($M{=}41.17, SD{=}9.49$) ($t(11){=}-3.55, p{=}0.0124$). Finally, users consistently awarded our full system the highest ratings across all five of our custom design-specific questions (Figure~\ref{fig:qualitative_additional}), with statistical significance (all with $p{<}0.05$) over the best baselines.
This demonstrates a clear subjective preference for correct detection results and ergonomic interaction provided by our full pipeline.

\paragraph{Participant comments}
After finishing the user study, we interviewed the participants to further obtain their comments.
Participant feedback revealed a strong preference for using physical objects as tangible interfaces, \revise{with users finding the interaction intuitive (\ie, easy to learn the mapping) and interesting.}
A key qualitative finding is that physical features—such as edges and protruding corners—serve as natural affordances that enable eyes-free localization (P1, P4, P8-P11). Notably, the interaction experience varied significantly across different form factors. Participants preferred distinct geometric elements mapped to specific tasks, such as using the tip of a dispenser for 0D discrete input, the rim of the pitcher lid for 1D continuous sliding, and broad caps of milk carton for 2D panning (P1). 
Despite the positive reception, participants highlighted tracking robustness as the primary limitation. Users reported occasional unintended inputs and tracking drift, particularly when interacting with the sugar dispenser (P1, P4, P5), suggesting future iterations to prioritize robustness against hand occlusion.

\paragraph{Discussion} In light of the results of our experiment, we can now attempt to answer the questions posed earlier.
\textbf{Q1:} 
Our full detection pipeline outperforms ablation baselines in terms of detection accuracy, with 100\% task completion rate (Table~\ref{tab:skip_test_statistics}) and high user preference (average score of $4.60$).
\textbf{Q2:}
Except for the 0D task, the improved detection of geometric elements brings better task performance.
This benefit is particularly prominent on the 2D task (38.48\% boost over the best baseline). 
The system usability is also validated by users with a substantially higher SUS score (+37.16\%) and lower workload (-33.20\%) than the best baseline.
\textbf{Q3:} 
In the simulated kitchen scenario, our system supports consistent, fluid microgesture interaction when users switch between different physical objects. 
On objects of different shapes and sizes, our full system \revise{enables} users to \revise{complete} scenario-specific tasks within a reasonable time (4.91 seconds for video time seeking and 7.53 seconds for adjusting the focal region), while achieving a small standard deviation of 0.88 and 0.59 seconds, respectively. 

A key takeaway of our work is that micro-geometry provides universal affordances: by mapping 0D, 1D, and 2D interactions to corners, edges, and surfaces, we bypass the need to understand object semantics. 
Furthermore, we highlight that while generative foundation models excel at extracting this geometry, they must be tightly coupled with human factors. 
By constraining vision algorithms with human reachability and ergonomic models, our framework ensures that detected results are not just visually accurate but physically usable. 
\revise{Finally, by scaling to everyday objects, we envision a future where users can seamlessly repurpose anything, anywhere and anytime, as a physical anchor for rapid control.}

\revise{

\subsection{User Study 2: Generalizability}
We further evaluate \ourname's generalizability across object and grasping variations on 10 everyday objects of different shapes and sizes: hand cream, spray bottle, clamp, coffee mug, book, scissors, fish can, milk carton, water pitcher, and sugar dispenser, each under two different grasping poses.
The same set of 12 participants from User Study 1 were asked to perform the same three tasks and we 
measured the mean task performance.
Across the expanded object set, the failed-trial rates are only 6.67\%, 1.67\%, and 2.43\% for the 0D, 1D, and 2D tasks, respectively.  Mean accuracy for the 0D task exceeds 98\% ($SD=0.34$ \%), with mean completion time 
$5.47$~s ($SD=0.99$~s) for the 1D task and
$9.81$~s ($SD=2.28$~s) for the 2D task, comparable to Section~\ref{subsubsec:result}.
While these results show encouraging evidence, full generality across all possible objects requires larger-scale and long-term validation, especially on boundary cases such as granular building blocks. 

\subsection{System-Level Evaluation}
At the system level, \ourname comprises three functional subsystems: hand-object tracking, geometric element detection, and real-time interaction. Below, we show evaluations on tracking robustness and detection accuracy; a detailed analysis of computational efficiency is provided in the supplementary material.

\paragraph{Tracking robustness to occlusion}
To examine the quality of object tracking, as external ground truth is difficult to obtain, we follow a stability-based protocol similar to~\cite{tangiar25xu}. That is, the object is kept physically stationary against a moving headset. We then measure deviations in the reported pose using mean absolute deviation (MAD) and standard deviation (SD). This process is repeated for three objects (sugar dispenser, milk carton, and water pitcher) in different sizes (small, medium, and large) and also for grasping poses with varying occlusion levels (0\%, 15\%, and 50\%) to study how hand occlusion affects tracking reliability. 
Tracking remains stable under moderate occlusion: position MAD/SD<3.0mm and rotation MAD/SD<1.5$^{\circ}$. Error increases at 50\% occlusion for smaller objects (5.5mm, 4.0$^{\circ}$). The pose deviation stays within our contact threshold for <50\% occlusion, which is sufficient in most cases.

\paragraph{Detection accuracy}
To examine how the pipeline identifies candidate interaction regions independent of downstream applications, we evaluate geometric-element detection accuracy across the expanded object sets by having two independent users assess whether each detected element in User Study 2 is correct. 
The accuracy for the candidates (before usability analysis) is around 62\% on average and increases to above 90\% for the final results (after usability analysis), which effectively reduces regions with low accessibility, insufficient geometric sharpness, or being too small.
}

\revise{
\section{Discussion, Limitations and Future Work}
\paragraph{Benefits of on-object microgestures}
\ourname targets immediate digital control while users hold task-relevant everyday objects, aligning with prior work on grasp-aware microgestures~\cite{sharma19,aponte24grav,caetano25graspr}.
Putting an object down to reach for a separate controller interrupts the physical task. \ourname instead turns accessible geometric features on the held object into controls, preserving the grasp and supporting multitasking.
A representative example is tutorial-guided cooking
(Figure~\ref{fig:application_scenarios}(c)), where users handle pans, containers, and ingredients while controlling an instructional video. \ourname enables pause, scrub, or zoom, with fast and fine-grained controls.
\paragraph{Comparison with alternative input modalities}
\ourname complements rather than replaces voice, gaze, and free-space gestures. Voice suits discrete and rough commands but is less effective for continuous and fine-grained control and less suitable in noisy or shared settings. Gaze-and-dwell enables rapid selection but occupies visual attention. Free-space gestures are expressive but may disrupt the working posture
and lack a physical reference for tactile feedback~\cite{he23ubiedge, jain23ubitouch, he24adaptui}. In contrast, \ourname provides fast and compact control while preserving the grasp.

\paragraph{Generative refinement failures}
In most cases, our generative 3D refinement improves geometric regularity around interaction regions, but it may over-simplify or distort the geometry if the relevant region is poorly observed or heavily occluded in the input view. 
Such a failure can affect the interaction accuracy due to strong misalignment. These cases can be detected by comparing the local geometry before and after refinement, for example, using the Chamfer distance over the candidate interaction region; large deviations indicate that the refined geometry may no longer preserve the physical interaction surface.

\paragraph{Limitations and future work}
First, the geometric-element detection pipeline, though being a one-time pre-processing, is not real time, primarily because fingertip-reachability modeling requires intensive self- and inter-collision calculations. We plan to accelerate these computations through GPU parallelization. Second, robustness partly depends on 3D object and hand tracking, which remains susceptible to occlusion and noise; three participants noted occasional instability. 
Leveraging world models for physical understanding may improve robustness.
Third, ATOM detects contacts based on the proximity between tracked fingertips and parameterized geometric elements, enabling generalization without per-object training. Learning-based alternatives could improve robustness after task-specific training on large object- and gesture-specific datasets. 
Fourth, ATOM focuses on single-finger input to maintain a clear, learnable vocabulary. A future extension is to support multi-finger and compound gestures by jointly tracking multiple fingertips and their interactions to allow simultaneous operation. 
Finally, ATOM emphasizes geometric affordance, since corners, edges, and surfaces are universal primitives that provide clear tactile feedback. 
Incorporating other physical attributes,~\eg, surface texture, as additional control dimensions could further improve discoverability, tactile feedback, and user preference.
}

\revise{
\section{Conclusion}
In this paper, we presented \ourname{}, a novel framework that transforms diverse everyday objects into tangible AR interfaces. By mapping fine-grained geometric elements — corners, edge segments, and surface patches — to 0D, 1D, and 2D microgestures, our approach enables intuitive digital control through the natural affordances of handheld objects.
To detect such small-scale and irregular features across diverse objects, we introduced a robust, fingertip-aware detection pipeline. Rather than relying on shape priors, \ourname{} uses the cross-domain capabilities of 2D and 3D generative foundation models to enhance and extract prominent geometry.
Furthermore, we incorporate human factors by 
formulating the usability of detected elements
to prioritize ergonomic and accessible interactions.
Through extensive user studies, we showed that \ourname{} surpassed ablation baselines in task performance, system usability, and cognitive workload,
with promising evidence of generalization to a broad object set.
These results demonstrate \ourname{}'s potential for repurposing everyday objects for 
rich, tangible micro-interactions, allowing users to maintain control while holding objects during 
daily activities.
}

\begin{acks}
\revise{
This work is supported in part by the Research Grants Council of the Hong Kong Special Administrative Region, China (Project No. T45-401/22-N) and the Start-up Grant from City University of Hong Kong (Project No. 9610677).
Yinqiao Wang acknowledges Xiaoyuan Zhou's assistance in producing the supplementary video.
}
\end{acks}

\bibliographystyle{ACM-Reference-Format}
\bibliography{review}

\end{document}


\title{Supplementary Material\\
\ourname: Geometry-Aware Microgesture \revise{towards} 
Object-Agnostic Tangible Interaction}


\renewcommand{\shortauthors}{Wang et al.}

\newcommand{\eg}{\emph{e.g.}\xspace} \newcommand{\Eg}{\emph{E.g.}\xspace}
\newcommand{\ie}{\emph{i.e.}\xspace} \newcommand{\Ie}{\emph{I.e.}\xspace}
\newcommand{\cf}{\emph{cf.}\xspace} \newcommand{\Cf}{\emph{Cf.}\xspace}
\newcommand{\etc}{\emph{etc.}\xspace} \newcommand{\vs}{\emph{vs.}\xspace}
\newcommand{\wrt}{w.r.t.\xspace} \newcommand{\dof}{d.o.f.\xspace}
\newcommand{\iid}{i.i.d.\xspace} \newcommand{\wolog}{w.l.o.g.\xspace}
\newcommand{\etal}{et al.\xspace}

\newcommand{\ourname}{ATOM\xspace}
\newcommand{\revise}[1]{#1}

\maketitle

\section{Usability Analysis of Fingertip Ergonomics}
\begin{figure}
    \includegraphics[width=\linewidth]{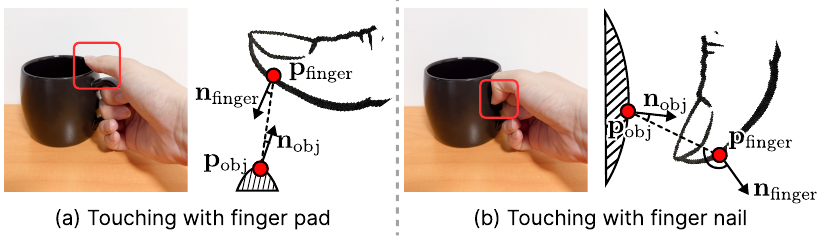}
    \caption{ Illustration of fingertip ergonomics. }
    \label{fig:supp_fingertip_erg}
\end{figure}
Figure~\ref{fig:supp_fingertip_erg} illustrates the impact of fingertip ergonomics used in our usability analysis.
Interactions feel natural when the finger pad is oriented toward the target region, such as when using the thumb to touch a mug rim (Figure~\ref{fig:supp_fingertip_erg} (a)).
Conversely, when the finger pad faces away from the surface—such as when touching the body of the mug (Figure~\ref{fig:supp_fingertip_erg} (b)), the interaction is typically uncomfortable, often forcing the user to slide using their fingernail.
To quantify this, we evaluate the alignment between the fingertip normal ($\mathbf{n}_{\text{finger}}$), the object's surface normal (per point, each denoted as $\mathbf{n}_{\text{obj}}$), and the connecting vector between $\mathbf{p}_{\text{finger}}$ and $\mathbf{p}_{\text{obj}}$. 
This allows us to prioritize candidate elements that are easily accessible by the finger pad (\eg, prioritizing the mug rim in Figure~\ref{fig:supp_fingertip_erg} (a) over the body area in Figure~\ref{fig:supp_fingertip_erg} (b)).

\section{Detailed Task Design in User \revise{Studies}}
\paragraph{0D task (pausing)}
In each trial of the 0D task, a tutorial video automatically plays from an initial timestamp, and the participant is asked to pause the playhead as close to a target timestamp as possible.
Across the four trials, the initial timestamps are set to $[0.3, 0.5, 0.7, 0.1]$ with corresponding target timestamps of $[0.4, 0.6, 0.8, 0.2]$ (in normalized coordinates). 
To simulate varying levels of challenge in the real world, the playback speed increases monotonically across the trials: $[\times1.0, \times2.0, \times3.0, \times4.0]$.

\paragraph{1D task (seeking)}
In the 1D task, the tutorial video remains paused. 
The participant is required to slide the playhead from an initial timestamp to a target timestamp. 
Across the four trials, the initial timestamps are $[0.0, 0.6, 0.4, 1.0]$, with corresponding target timestamps of $[0.2, 0.4, 0.6, 0.8]$.

\paragraph{2D task (panning)}
In the 2D task, the video is paused and zoomed in ($\times 2.0$).
The participant must drag the focus point from the screen center (\ie, $(0.5, 0.5)$ in normalized coordinates) to a specified target point in the video content.
Across the four trials, the target points are $[(0.25, 0.25),(0.25, 0.75),(0.75, 0.75),(0.75, 0.25)]$.

For both the 1D and 2D tasks, the dwell time required to successfully complete a trial is set to one second.

\section{Implementation Details}
\paragraph{Fingertip reachability modeling}
\begin{algorithm}
\caption{Fingertip Reachability Modeling}
\label{alg:finger_space}
\KwIn{Initial hand parameters $P_{base}$, hand shape $S$, target finger $f$, sample steps $N$, object geometry $O$}
\KwOut{Valid fingertip positions $V_{tips}$}

$R_{prox\_spread}, R_{prox\_bend}, R_{inter\_bend}, R_{dist\_bend}$$ \gets \text{GetJointMotionRanges}(f)$\;

$\Theta_{spread}, \Theta_{pbend}, \Theta_{ibend}, \Theta_{dbend} \gets \text{Discretize}(R_{prox\_spread}, R_{prox\_bend}, R_{inter\_bend}, R_{dist\_bend}, N)$\;
$\Omega \gets \Theta_{spread} \times \Theta_{pbend} \times \Theta_{ibend} \times \Theta_{dbend}$\;
$V_{tips} \gets \text{EmptyArray}(N^4)$\;

\ForEach{pose $\theta \in \Omega$ in batches}{
    $P_{temp} \gets \text{UpdateFingerPose}(P_{base}, f, \theta)$\;
    $\mathcal{H}_{mesh}, \mathcal{H}_{joints} \gets \text{ForwardKinematics}(P_{temp}, S)$\;

    \If {\textbf{not}\ \text{HasSelfPenetration}($\mathcal{H}_{mesh})$ \textbf{and} \textbf{not}\ \text{HasObjectPenetration}($\mathcal{H}_{mesh}, O)$}
    {
        $V_{tips}[\theta] \gets \text{GetFingertipPosition}(\mathcal{H}_{joints}, f)$\;
    }
}

\Return{$V_{tips}$}\;
\end{algorithm}
The pseudo code for modeling fingertip reachability is provided in Algorithm~\ref{alg:finger_space}. 
To analyze the movement space of the active finger, we first generate all possible hand poses where only that specific finger moves, constrained by its anatomical range of motion~\cite{nielsen04}. 
These hand poses are then converted into 3D hand meshes and joint coordinates using the MANO model~\cite{embodied17}.
Finally, to ensure the physical plausibility of the reachable space, we filter out any poses that exhibit self-intersection or collisions with the object.
The fingertip positions from the valid poses are then voxelized to form the reachable region.

\paragraph{Corner consolidation}
While the Harris's detector~\cite{Harris1988ACC} yields relatively accurate results on the refined local edge map, it often detects dense clusters of points within a single corner region. 
To enable precise interaction with true corners, we adopt DBSCAN~\cite{dbscan96} to adaptively group these dense points into distinct clusters. 
We then extract the centroid of each cluster as a candidate corner. 

\paragraph{Geometric element projection}
To support 3D interaction, we finally back-project the detected 2D corners, edge segments, and surface patches within the local area (Section 4.2 of the main paper) into 3D space.
This is achieved using the object's rendered depth map and the egocentric camera parameters provided by the head-mounted display.

\paragraph{Hyperparameters}
During element prioritization, we establish several hard thresholds for filtering and processing. 
Specifically, we discard a candidate element if any of the following criteria are met:
(i) over 60\% of its emitted rays are blocked by the object during accessibility testing;
(ii) the average angle between its normal vector and those of its nearest neighbors is less than $30^\circ$ during sharpness testing;
(iii) it is an edge shorter than 1~cm or a surface with an area smaller than 4~cm$^2$.
Besides, we filter out any points on the element (iv) that are located within 1~cm of other candidates.

For ranking, we empirically set the weights of the usability factors as follows: $w_{\text{rch}}=3.0$, $w_{\text{acc}}=1.0$, $w_{\text{erg}}=1.0$, $w_{\text{abg}}=0.5$, and $w_{\text{shp}}=3.0$.
The larger weight assigned to reachability stems from preliminary experiments, where we observed that users prefer accessing target regions requiring minimal finger movement.
Additionally, a larger weight for sharpness ensures that selected corners and edges are geometrically distinct from flat surfaces.
Conversely, we assign a lower weight to gesture ambiguity because it often conflicts with reachability; penalizing ambiguity pushes selected elements further apart, which might make them more difficult to reach. 
Furthermore, since we already filter out points located in close proximity to other geometric elements prior to ranking, the remaining candidate elements are already sufficiently separated to support fluid interaction.

\revise{
\paragraph{Setup requirements}
The current deployment of \ourname comprises three stages, separated according to their computational efficiency and execution frequency.
(i) \textit{Object preparation (offline)}: Each object is scanned once using multi-view RGB-D input, specifically RealityKit Object Capture~\cite{apple_object_capture_2026}, to reconstruct its geometry, with no prior template required.
During scanning, the object should remain visible and static; otherwise, the hand may occlude surfaces and prevent them from being captured.
The features of the reconstructed mesh are then extracted for ARKit~\cite{apple_arkit_2024} object tracking.
(ii) \textit{Geometric element detection (offline)}: After ARKit begins tracking the object, the user can hold it in an arbitrary grasping pose. For each pose, the egocentric-view RGB image and pre-scanned mesh are processed by our detection pipeline (Figure~3 of the main paper). This stage currently runs offline because of its computational cost and reliance on external API calls.
(iii) \textit{Real-time interaction (online)}: The detected elements are cached on the device and reused for subsequent interactions with the same object and grasping pose. During the online loop, the system combines the tracked object pose and hand pose to perform microgestures in real time.

\section{Efficiency analysis}
We analyze per-stage timing as below. The offline processes include: (i) per-object: pre-scanning/feature extraction ($\sim$10h) and 3D enhancement ($\sim$5min), which are cached once; (ii) per-pose: 2D refinement ($\sim$20s), element detection ($\sim$2min), and usability analysis ($\sim$10s) — cached after first run. 
Once a pose is processed, its detection results are cached on-device, allowing users to pick up any previously scanned object and immediately interact with AR content.
The online interaction loop runs at $\sim$82 FPS on Apple Vision Pro (tracking $\sim$1.5ms, contact $\sim$0.2ms, response $\sim$0.01ms). At use time, only the initial detection for each new pose is performed off-device. 
In future deployments, we plan to execute these currently offline processes on-device using lightweight models and GPU acceleration.
}

\bibliographystyle{ACM-Reference-Format}
\bibliography{review}